\documentclass[%
 reprint,
superscriptaddress,
 amsmath,amssymb,
 aps,
 prl,
]{revtex4-2}
\usepackage{graphicx}
\usepackage{dcolumn}
\usepackage{bm}

\usepackage{color}
\usepackage{braket}
\usepackage{hyperref}

\newcommand*{\Tr}{\text{Tr}}

\begin{document}
\title{Non-Hermitian Hamiltonians Violate the Eigenstate Thermalization Hypothesis}

\author{Giorgio Cipolloni}
\email{gc4233@princeton.edu}
 \affiliation{Princeton Center for Theoretical Science, Princeton University, Princeton, NJ 08544, USA}
\author{Jonah Kudler-Flam}%
 \email{jkudlerflam@ias.edu}
 \affiliation{Princeton Center for Theoretical Science, Princeton University, Princeton, NJ 08544, USA}
 \affiliation{School of Natural Sciences, Institute for Advanced Study, Princeton, NJ 08540 USA}

\date{\today}

\begin{abstract}
The Eigenstate Thermalization Hypothesis (ETH) represents a cornerstone in the theoretical understanding of the emergence of thermal behavior in closed quantum systems. The ETH asserts that expectation values of simple observables in energy eigenstates are accurately described by smooth functions of the thermodynamic parameters, with fluctuations and off-diagonal matrix elements exponentially suppressed in the entropy. We investigate to what extent the ETH holds in non-Hermitian many-body systems and come to the surprising conclusion that the fluctuations between eigenstates is of equal order to the average, indicating no thermalization. We support this conclusion with mathematically rigorous results in the Ginibre ensemble and numerical results in other ensembles, including the non-Hermitian Sachdev-Ye-Kitaev model, indicating universality in chaotic non-Hermitian quantum systems.
\end{abstract}

\maketitle

\textit{Introduction.}---Thermalization in quantum systems describes the dynamical process of out-of-equilibrium states settling down to quasi-equilibrium states that are well-described by quantum statistical mechanics and thermodynamics. Understanding which systems thermalize and how they do so is a foundational question in quantum physics that has received tremendous attention in the past decades \cite{2018RPPh...81h2001D}.

In recent years, the field of \textit{non}-Hermitian quantum physics has emerged as a generalization of the standard paradigm of Hermitian physics to describe systems with dissipation \cite{2020AdPhy..69..249A}. It is natural to ask if and how thermalization manifests in these systems. This direction goes under the name of \textit{dissipative quantum chaos} and has recently enjoyed several interesting results \cite{2019PhRvL.123i0603H,2019PhRvL.123y4101A,2020PhRvX..10b1019S, 2021PhRvL.127q0602L,2021arXiv211212109S,2021arXiv211213489K,2019PhRvL.123n0403D,2020PhRvL.124j0604W,2021PhRvR...3b3190S,2023PhRvL.130a0401C,2022arXiv221200605K,2022arXiv221004093K,2022PhRvR...4d3196X,2022arXiv221200474S,2022arXiv221101650G,2022arXiv221007959C,2022arXiv221001695G}. 

The Eigenstate Thermalization Hypothesis (ETH) is the paradigmatic description or definition of thermality in closed quantum systems. The ETH asserts that for chaotic quantum systems, the matrix elements of simple operators are smoothly varying on the diagonal in the energy eigenbasis $\{\ket{E_i} \}$ with entropically suppressed fluctuations
\begin{align}
    \label{ETH_eq}
    \bra{ E_i} \mathcal{O} \ket{E_j} = f_{\mathcal{O}}(\overline{E})\delta_{ij} + e^{-S(\overline{E})/2}\omega_{\mathcal{O}}\left(\overline{E},\Delta E\right)R_{ij}.
\end{align}
Here, $f_{\mathcal{O}}$ and $\omega_{\mathcal{O}}$ are smooth $O(1)$ functions, $\overline{E} = \frac{E_i + E_j}{2}$, $\Delta E = |E_i - E_j|$,  and $R_{ij}$ are pseudorandom numbers with unit variance. This describes thermalization because it implies that expectation values of simple operators can be accurately approximated using only the thermodynamic parameters such as energy. This statement applies to late-time states following time evolution and the energy eigenstates themselves.

\begin{figure}
    \centering
    \includegraphics[width = .48\textwidth]{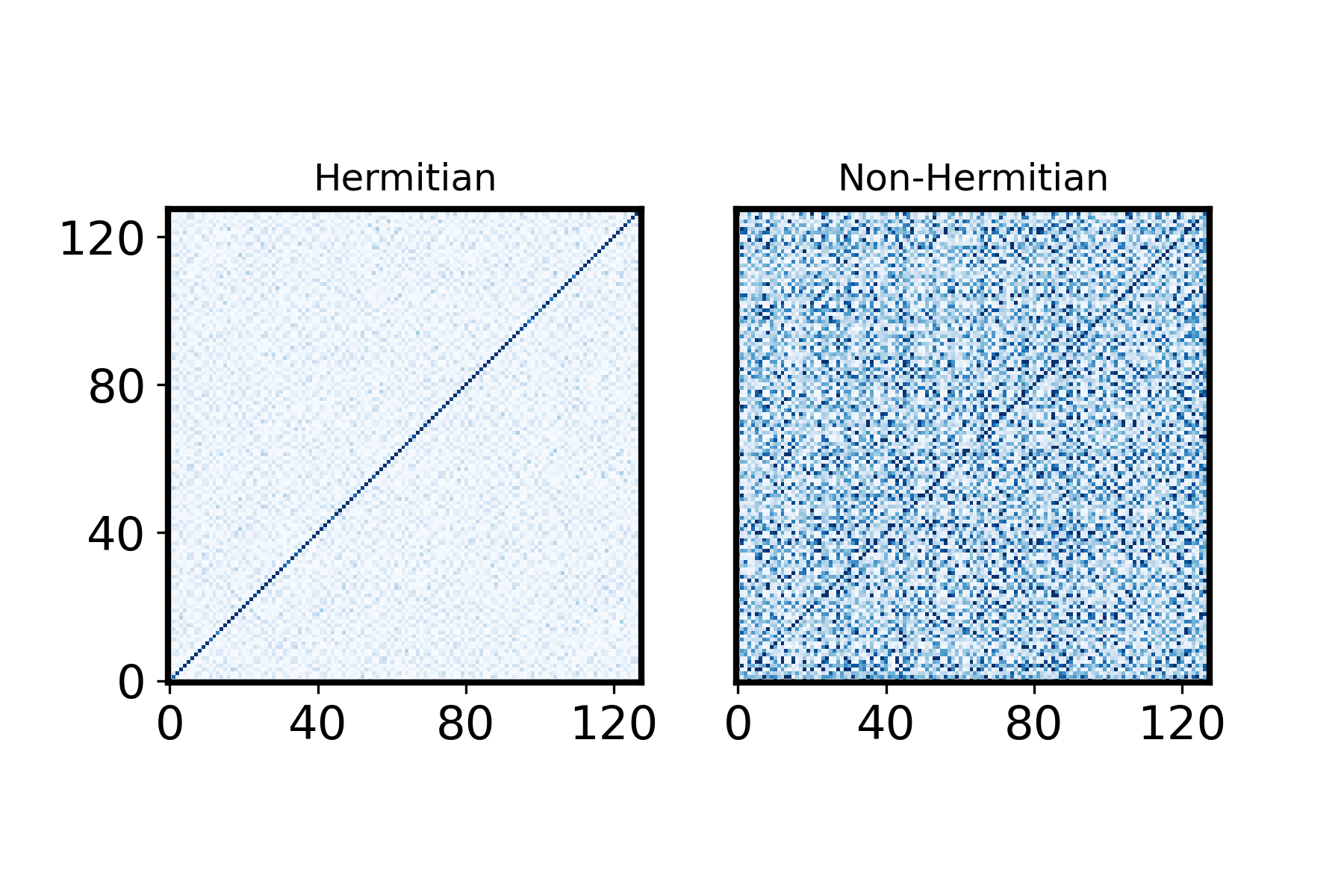}
    \caption{The absolute value of the matrix elements of the single-site number operators, $\hat{n}_{ij} := |\bra{ L_i } \hat{n} \ket{R_j}|$, are shown for a single realization of the Hermitian SYK model (left) and non-Hermitian SYK model (right). The x and y-axes label the left and right eigenvectors respectively. While $\hat{n}_{ij} $ is almost exclusively supported on the diagonal in the Hermitian case, there are large fluctuations across the entire matrix in the non-Hermitian case.}
    \label{noETH_SYK}
\end{figure}

In this Letter, we ask if there is an analog of \eqref{ETH_eq} for non-Hermitian Hamiltonians that characterizes dissipative quantum chaos. Non-Hermitian operators generally have distinct left and right eigenvectors, $\{\bra{L_i}\}$ and $\{\ket{R_i}\}$, that form a biorthonormal basis
\begin{align}
    \bra{L_i}R_j\rangle = \delta_{ij}.
    \label{biorth_eq}
\end{align}
Surprisingly, we observe that the eigenstates of the non-Hermitian version of the Sachdev-Ye-Kitaev (SYK) model \cite{2020PhRvR...2b3286H}, the canonical model of many-body quantum chaos \cite{1993PhRvL..70.3339S, Kitaev2015,2016PhRvD..94j6002M}, do \textit{not} thermalize; the fluctuations, both on and off the diagonal, are just as large as the mean for expectation values of simple operators. The stark contrast between matrix elements for the Hermitian and non-Hermitian SYK models is shown in Figure \ref{noETH_SYK}. The $N$ Majorana fermion Hamiltonians are of the form
\begin{align}
  H_{nSYK} = \sum^N_{i_1 < i_2 <\dots<i_q}\left(J_{i_1i_2\dots i_q}+iM_{i_1i_2\dots i_q} \right)\psi_{i_1}\psi_{i_2}\dots \psi_{i_q},
\end{align}
where $J_{i_1i_2\dots i_q}$ and $M_{i_1i_2\dots i_q} $ are i.i.d.~real Gaussian random variables with zero mean, variance $\frac{2}{N^{q-1}}$, and $\{\psi_i, \psi_j \} = 2\delta_{ij}$. $M_{i_1i_2\dots i_q}  = 0$ in the Hermitian case.

We will explain this striking phenomenon analytically by modeling non-Hermitian quantum chaotic Hamiltonians by random matrices. The connection between chaotic quantum systems and random matrix theory for both Hermitian and non-Hermitian systems is well-known \cite{bohigas1984characterization, berry1977level, PhysRevLett.61.1899,PhysRevLett.62.2893}
and provides a simple way to motivate \eqref{ETH_eq}. 
Indeed, it is widely believed that non-integrable quantum systems exhibit the same spectral statistics as Hermitian random matrices. In the context of random Hamiltonians, ``simple'' operators are deterministic operators, independently chosen from the Hamiltonian. For $N\times N$ Wigner matrices, \eqref{ETH_eq} has been proven rigorously \cite{2021CMaPh.388.1005C, cipolloni2023gaussian}.

In order to model typical chaotic non-Hermitian Hamiltonians, we study the eigenvectors of matrices drawn from the Ginibre ensemble defined by matrices with independent and identically distributed (i.i.d.) complex Gaussian matrix elements \cite{1965JMP.....6..440G}. We compute the behavior of the matrix elements of deterministic operators, precisely characterizing the large fluctuations. The new phenomena, not seen in the Hermitian case, include the aforementioned large fluctuations as well as correlations between matrix elements.
Moreover, we demonstrate that our rigorous results for the Ginibre ensemble are \textit{universal}, with extended regimes of validity in highly non-Gaussian ensembles such as the complex Bernoulli and uniform ensembles.

\textit{Structure of Ginibre Eigenvectors.}---
Let $H$ be an $N\times N$ matrix drawn from the complex Ginibre ensemble. We denote its eigenvalues by $\lambda_1,\dots, \lambda_N$ and the corresponding biorthonormal left and right eigenvectors by $\bra{L_i},\ket{R_i}$.
The distribution of $H$ is invariant under unitary conjugation. Following \cite{chalker1998eigenvector, mehlig2000statistical, mehlig1998eigenvector}, we write $H=UTU^{\dagger}$, with $U$ independent of $T$ and uniformly distributed on the unitary group, and
\begin{align}
    T:=\left(\begin{matrix}
\lambda_1 & T_{12} &\dots & T_{1N} \\
0 & \lambda_2 & \dots & T_{2N} \\
\vdots & \ddots & \ddots & \vdots \\
0 & \dots & 0 & \lambda_N
\end{matrix}\right),
\end{align}
where $T_{ij}$ are uncorrelated Gaussian random variables. The eigenvectors of $T$ are of the form
\begin{equation}
    \label{eq:uschange}
\begin{split}
\ket{\widetilde{R}_1}&=(1,0\dots,0)^T, \qquad \ket{\widetilde{R}_2}=(a,1,0\dots, 0)^T \\
\ket{\widetilde{L}_1}&=(b_1,\dots, b_N)^T, \qquad \ket{\widetilde{L}_2}=(d_1,\dots,d_N)^T,
\end{split}
\end{equation}
with $b_1=1$, $d_1=0$, $d_2=1$, $a=-{b_2}^*$.

\textit{Eigenvector overlaps.}--- A fundamental quantity in the analysis of the spectrum of $H$ are the so called eigenvector overlaps
\begin{equation}
    O_{ij}:= \bra{R_j}R_i\rangle\bra{L_i}L_j\rangle.
\end{equation}
In contrast to the eigenvectors of Hermitian Hamiltonians, where $\bra{R_i}R_j\rangle = \bra{L_i}L_j\rangle = \delta_{ij}$, the eigenvector overlaps can be quite large and highly correlated. 

We will only need to consider overlaps involving one or two eigenvectors, which will be denoted by $O_{11},O_{12},O_{22} $, i.e.~we drop the $i,j$--dependence. These overlaps can be expressed in terms of the quantities appearing in \eqref{eq:uschange} as
\begin{align}
\begin{aligned}
        O_{11} = \sum_{i = 1}^N |b_i|^2,
    \quad
    O_{12} = -b_2^*\sum_{i = 2}^N b_id_i^*,
    \\
    O_{22} = (1+ |b_2|^2)\sum_{i = 2}^N |d_i|^2.
\end{aligned}
\end{align}
Critical to our analysis are Theorems 1.1 and 1.4 of \cite{bourgade2020distribution}, where the overlaps are evaluated at large $N$ (see also \cite{fyodorov2018statistics} for the diagonal overlaps). The full distribution of the diagonal overlaps converge to
\begin{align}
\label{O11_distr}
    O_{11} \to
    \frac{N(1-|\lambda_1|^2)}{\gamma_2},
\end{align}
with $\gamma_2$ being a Gamma random variable with density $xe^{-x}$. The other overlaps that we will make use of are only known in expectation and in second moment
\begin{align}
\mathbb{E} O_{12}&\rightarrow -\frac{1-\lambda_1{\lambda_2}^*}{N|\lambda_1-\lambda_2|^4}, \\
\mathbb{E} |O_{12}|^2&\rightarrow\frac{(1-|\lambda_1|^2)(1-|\lambda_2|^2)}{|\lambda_1-\lambda_2|^4},\label{O12O12}
\\
\mathbb{E} O_{11}O_{22}&\rightarrow \frac{(1-|\lambda_1|^2)(1-|\lambda_2|^2)}{|\lambda_1-\lambda_2|^4} \label{O11O22}
\\
\nonumber
&\times \frac{1+N^2|\lambda_1-\lambda_2|^4-e^{-N|\lambda_1-\lambda_2|^2}}{1-e^{-N|\lambda_1-\lambda_2|^2}}.
\end{align}

To study the ETH we consider $\bra{L_i} A \ket{R_j}$, and their correlations. In particular, we will compute the expectation and the variance for these quantities using Weingarten calculus, which provides closed form expressions for integrals of the unitary group with respect to the Haar measure \cite{2002math.ph...5010C}:
\begin{align}
    &\int \left[ dU \right] U_{i_1, j_1} \dots U_{i_n, j_n}U^*_{i_1', j_1'}\dots U^*_{i_n', j_n'}
    \nonumber 
    \\
    &= \sum_{\sigma, \tau \in S_n} \delta_{i_1i_{\sigma(1)}'}\dots\delta_{i_ni_{\sigma(n)}'}\delta_{j_1j_{\tau(1)}'}\dots\delta_{j_nj_{\tau(n)}'}\mbox{Wg}(N, \sigma \tau^{-1}).
\end{align}
Here, $\text{Wg}$ are the so-called Weingarten functions, whose explicit form may be written in terms of the characters of the symmetric group. For our purposes, we will only need up to the fourth moment of the unitary group. We find the \verb|Mathematica| package \verb|RTNI| \cite{2019JPhA...52P5303F} to be helpful in explicit evaluations of the 576 terms in the sum. We will only present the leading order (in $N$) results.

Importantly, the expectations we compute are only with respect to the $U$'s. The independent random variables in the $T$ matrix are not averaged over, leaving a resulting distribution that aids in conjectures.

\begin{figure}
    \centering
    \includegraphics[width = .45\textwidth]{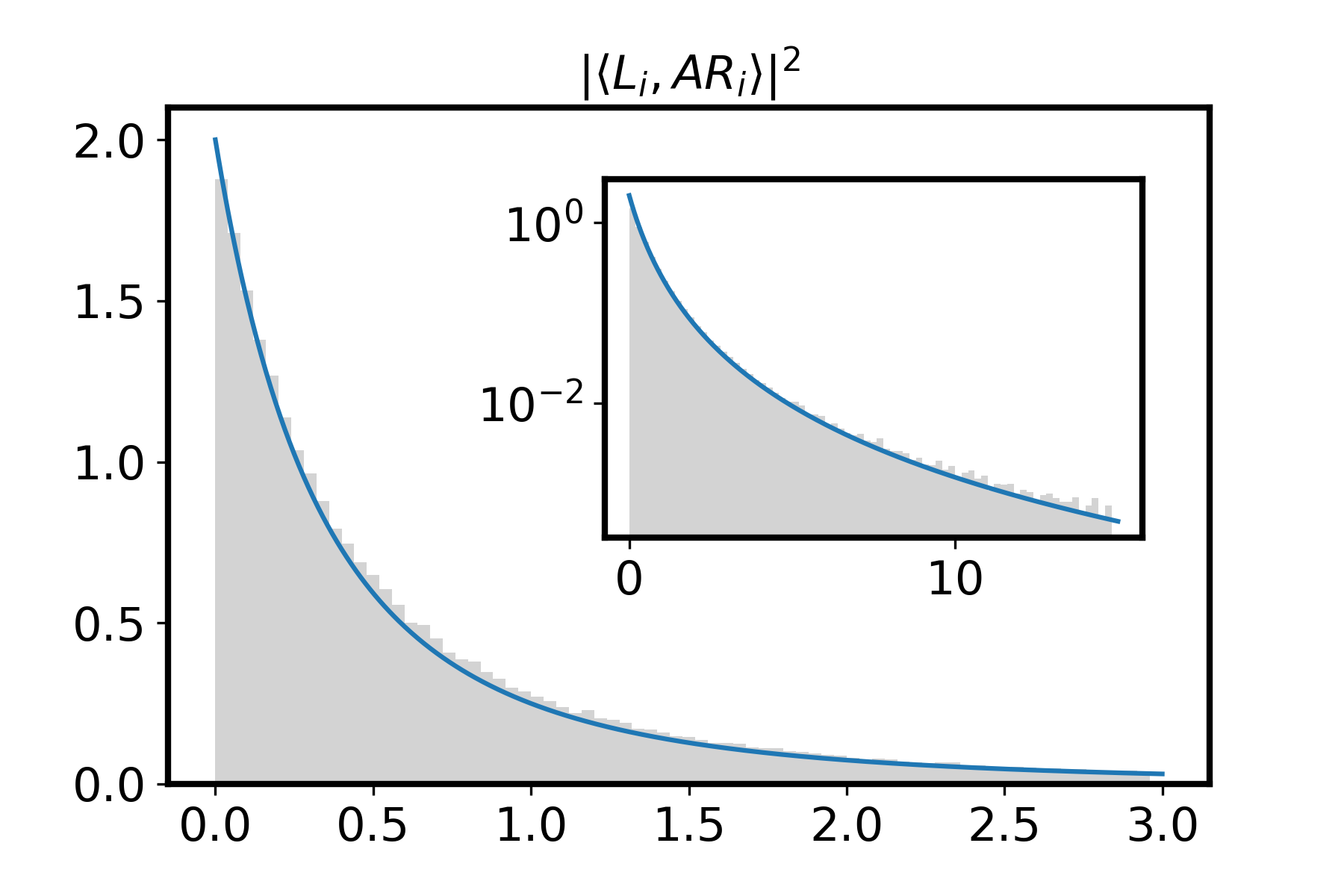}
    \caption{The absolute value squared of the matrix elements weighted as in the left hand side of \eqref{conj_distr} of a deterministic matrix. We take $1000$ realizations for $N = 200$. The blue line is \eqref{conj_curve}. The inset is the same data but with logarithmic scaling of the y-axis.}
    \label{diag_check_nh_eth}
\end{figure}

\textit{Diagonal ETH.}---We first consider the expectation values of operators i.e.~the diagonal portion of ETH. The Haar integral is straightforward, giving the normalized trace of the observable
\begin{align}
    \mathbb{E}_U \bra{L_i} A \ket{R_i} = \frac{\Tr \left(A\right)}{N}.
\end{align}
To estimate the average size, we instead need the square
\begin{align}
    \begin{aligned}
    \label{diag_fluct_eq}
       \mathbb{E}_U  |\bra{L_i} A \ket{R_i}|^2 = \frac{1}{N^2}\left|\Tr\left(A \right)\right|^2
    \\
    +\frac{O_{11}}{N^2 }\left(\Tr\left(A^{\dagger} A \right)-N^{-1}\left|\Tr \left(A\right)\right|^2 \right).
\end{aligned}
\end{align}
From \eqref{O11_distr} we see that the second line is the same order as the first, so the fluctuations in the expectation values are not suppressed, strongly violating \eqref{ETH_eq}.

We {may} expect that at large-$N$ {(for a motivation see the \emph{Universality} section below)}
\begin{equation}
   \label{conj_distr}
\frac{N}{(1-|z|^2)\Tr\left(A^{\dagger} A \right)} \left| \bra{L_i} A \ket{R_i} -N^{-1}\Tr (A)\right|^2 \to  \frac{|\xi|^2}{ \gamma_2},
\end{equation}
in distribution, where $\xi$ a complex Gaussian random variable independent of $\gamma_2$. The probability density function is
\begin{align}
\label{conj_curve}
    p(x) = \frac{2}{(1+ x)^{3}}.
\end{align}
Due to the $\gamma_2$ in the denominator, there is heavy tail in this distribution, a characteristic of non-Hermitian systems, also seen in the entanglement spectrum of typical states \cite{2023PhRvL.130a0401C}, which is in stark contrast with the Hermitian case. 
We verify this expectation numerically in Figure \ref{diag_check_nh_eth}.

\textit{Off-Diagonal ETH.}---Off of the diagonal, the ETH implies that the matrix elements are exponentially suppressed. Because there is a scaling ambiguity in \eqref{biorth_eq} taking $\ket{R_i} \rightarrow c_i\ket{R_i}$ and $\bra{L_i} \rightarrow c_i^{-1}\bra{L_i}$, a well-defined (scale independent) notion of the off-diagonal elements is $\bra{L_i}A\ket{R_j}\bra{L_j}A\ket{R_i}$. Averaging over unitaries, we find (for $i \neq j$)
\begin{align}
\begin{aligned}
        \mathbb{E}_U \bra{L_i} A \ket{R_j}{\bra{L_j} A \ket{R_i}}
        \\
        = \frac{1}{N^2}\left(\text{Tr}\left(A^2\right)-N^{-1}{\text{Tr}\left(A\right)^2}\right).
\end{aligned}
\end{align}
There is no $O_{12}$ dependence because we have not taken the complex conjugate of the second term. While the  above expression is suppressed in $N$, the average size is large which can be seen from the absolute value squared
\begin{align}
\begin{aligned}
        \mathbb{E}_U |\bra{L_i} A \ket{R_j}{\bra{L_j} A \ket{R_i}}|^2
        \\
        = \frac{O_{11} O_{22} \text{Tr}[A^{\dagger}A]^2+|O_{12}|^2 \text{Tr}[A^{\dagger}A]^2}{N^4}
\end{aligned}
\end{align}
From \eqref{O11O22}, we see that this is $O(1)$, so there is no entropic suppression off the diagonal, our second exhibition of a strong violation of the ETH. The second term is only the same order as the first when the eigenvalues are very close.

\textit{Correlation for nearby energies.}---The final novel feature in non-Hermitian Hamiltonians that we explore is the large correlation between nearby eigenvalues. This means that even though there are large fluctuations in the diagonal elements (see \eqref{diag_fluct_eq}), these fluctuations are correlated. For simplicity, we present the answer for traceless observables
\begin{align}
\begin{aligned}
        \mathbb{E}_U |\bra{L_i} A \ket{R_i}{\bra{L_j} A \ket{R_j}}|^2 \\-  \mathbb{E}_U |\bra{L_i} A \ket{R_i}|^2\mathbb{E}_U|{\bra{L_j} A \ket{R_j}}|^2
        = 
         \frac{\left|O_{12}\right|^2
   \text{Tr}[A^{\dagger} A]^2}{N^4}.
\end{aligned}
\end{align}
As seen from \eqref{O12O12}, the correlation is large when the eigenvalues are very close ($\lambda_i -\lambda_j \sim N^{-1/2}$) but quickly decays as the eigenvalues are separated.

\textit{Universality.}---We expect that the results that we have derived for Ginibre matrices are universal. In particular, if we consider a matrix $H$ with i.i.d.~entries but not necessarily with Gaussian distribution (or even with some specific correaltion structure), then the same convergence results should hold.

To see this we rely on the \emph{Hermitization trick} from \cite{1997NuPhB.504..579F} and used in \cite{2023PhRvL.130a0401C}. The following discussion is similar to the one presented in \cite{2023PhRvL.130a0401C}, but we repeat it here for the reader's convenience. More precisely, we define the Hermitian matrix \begin{equation}
\label{eq:hemr}
H^z:=\left(\begin{matrix}
0 & H-z \\
(H-z)^{\dagger} & 0
\end{matrix}\right),
\end{equation}
with $z\in\mathbb{C}$. Observing that
\begin{equation}
\lambda\in \mathrm{Spec}(H)\Leftrightarrow 0\in \mathrm{Spec}(H^\lambda),
\end{equation}
one can use $H^z$ to study spectral properties of $H$ itself when $z$ is very close to one of its eigenvalues. We point out that $H^z$ satisfies a chiral symmetry which induces a symmetric spectrum about zero. As a consequence, the eigenvectors $\ket{{\bf w}_{\pm i}^z}$ are of the form $\ket{{\bf w}_{\pm i}^z}=(\ket{{\bf u}_i^z},\pm\ket{{\bf v}_i^z})$, with $\ket{{\bf u}_i^z},\ket{{\bf v}_i^z}\in\mathbb{C}^N$. Here $\ket{{\bf u}_i^z},\ket{{\bf v}_i^z}$ denote the  left and right singular vectors of $H-z$, i.e.
\begin{equation}
\label{eq:defsingv}
(H-z)\ket{{\bf v}_i^z}=E_i^z\ket{{\bf u}_i^z}, \quad
(H-z)^{\dagger}\ket{{\bf u}_i^z}=E_i^z\ket{{\bf v}_i^z},
\end{equation}
with $E_i^z\ge 0$ the corresponding singular values. Since the singular values coincide with the eigenvalues of $H^z$ (in absolute value), these two representations are equivalent. We choose to present both since in the following discussion may be more convenient to refer to one or the other.

\begin{figure}
    \centering
    \includegraphics[width = .4 \textwidth]{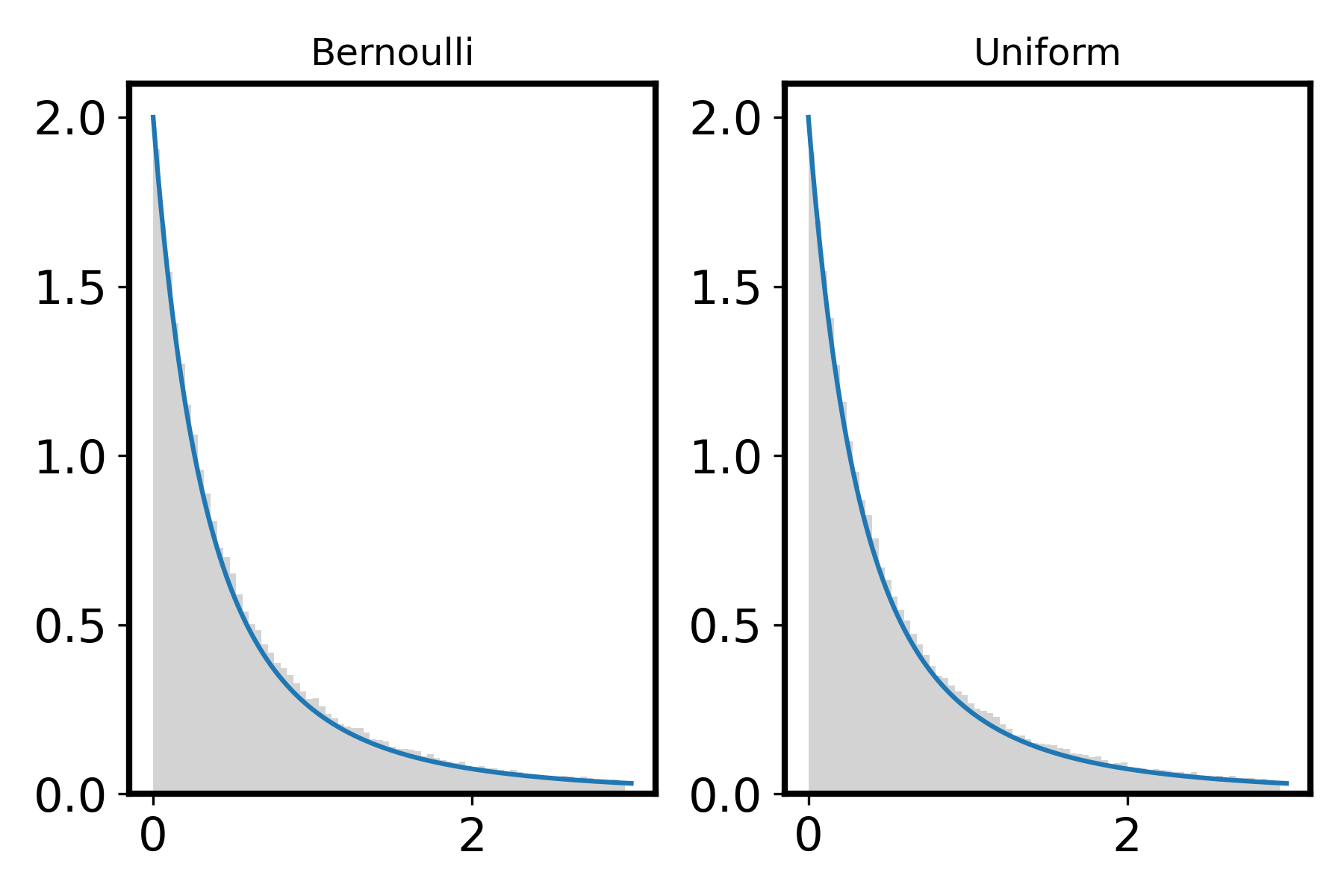}
    \caption{The absolute value squared of the matrix elements weighted as in \eqref{conj_distr} are shown for the complex Bernoulli and random complex uniform matrices. The blue line is \eqref{conj_curve}, motivating universality. We take $1000$ realizations for $N = 200$ and a fixed deterministic matrix $A$.}
    \label{fig:universality}
\end{figure}

Using the relation \eqref{eq:defsingv} and the definition of left and right eigenvectors $\bra{L_i}, \ket{R_i}$, we write
\begin{equation}
\label{eq:hermnonherm}
     \bra{L_i} A \ket{R_i}=\frac{{\bra{{\bf u}}A\ket{\bf v}}}{{\bra{{\bf u}} {\bf v}\rangle}},
\end{equation}
Here $\ket{\bf u}=\ket{{\bf u}_1^{\lambda_i}}$, $\ket{\bf v}=\ket{{\bf v}_1^{\lambda_i}}$. The key feature of the equality \eqref{eq:hermnonherm} is that we can express the non-Hermitian quantity in the LHS in terms of Hermitian singular vectors, which are much better understood. For fixed $z$, the distribution of $\ket{{\bf u}_i^z}, \ket{{\bf v}_i^z}$ can be understood using the Dyson Brownian motion (DBM) for eigenvectors introduced in \cite{2013arXiv1312.1301B} (see also \cite{2020arXiv200508425M,2022CMaPh.391..401B, benigni2022fluctuations, 2021arXiv210306730C,2022arXiv220301861C,  bourgade2018random, cipolloni2023gaussian, benigni2022fluctuations} for a detailed explanation of the DBM analysis in the mathematics literature). More precisely, in order to apply DBM, for fixed $z$ we write
\begin{equation}
\label{eq:quadform}
\bra{{\bf u}_1^z}A\ket{{\bf v}_1^z}=\bra{{\bf w}_1^z}\mathcal{A}\ket{{\bf w}_1^z}, \qquad \mathcal{A}:=\left(\begin{matrix}
0 & A \\
0 & 0
\end{matrix}\right),
\end{equation}
and notice that \eqref{eq:quadform} consists of a quadratic form with Hermitian eigenvectors $\ket{{\bf w}_1^z}$ for which the Hermitian DBM techniques from \cite{2013arXiv1312.1301B, 2020arXiv200508425M,2022CMaPh.391..401B, benigni2022fluctuations, 2021arXiv210306730C,2022arXiv220301861C, bourgade2018random, cipolloni2023gaussian} can apply. We defer the interest reader to \cite[Supplemental Material]{2023PhRvL.130a0401C} for a gentle introduction of the DBM analysis for Hermitian eigenvectors. {In particular, this analysis gives an insight to motivate the convergence in \eqref{conj_distr}.} However, here there is a caveat; the equality in \eqref{eq:hermnonherm} holds only for $z=\lambda_i$ in \eqref{eq:defsingv}, i.e. when $z$ is not fixed but rather equal to an eigenvalue. Even though computing the distribution of \eqref{eq:quadform} for fixed $z$ is not enough to compute the distribution of the LHS of \eqref{eq:hermnonherm}, we expect that this gives the right answer. Indeed, we also confirm this universality phenomenon numerically in Figure \ref{fig:universality} where we consider two non-Hermitian random matrix ensembles obeying the circular law (eigenvalues uniformly distributed in unit circle). These are the complex Bernoulli ensemble (entries are independent $\pm$ real and imaginary parts) and random complex uniform ensemble (entries are uniformly drawn from the complex unit circle).  

Returning to the non-Hermitian SYK model, we note that all of the qualitative phenomena observed in Figure \ref{noETH_SYK} have now been explained from the Ginibre ensemble. However, we do not expect the equations for the Ginibre ensemble to quantitatively agree with the SYK model because the eigenvalues of the SYK model do not obey the circular law. We do still expect that, after performing the proper $z$--rescaling the LHS of \eqref{conj_distr}, the expectation values of operators are distributed as in \eqref{conj_curve}. This is analogous to the Hermitian case where analytical results in the Gaussian ensembles only qualitatively agree with chaotic Hamiltonians because the eigenvalues of generic Hamiltonians do not obey the semi-circle law.

\textit{Discussion.}---In this Letter, we have shown that the Eigenstate Thermalization Hypothesis is strongly violated in non-Hermitian many-body systems. While initially observed in the SYK model, we gained an analytical understanding of this general phenomenon by modeling quantum chaotic non-Hermitian Hamiltonians using random matrix theory. We rigorously proved various novel features of the matrix elements of simple operators. These include (1) large fluctuations along the diagonal in the energy eigenbasis, of the same order as the mean, (2) fluctuations off of the diagonal that were just as large as those on the diagonal, and (3) strong correlations between matrix elements with nearby energies. None of these phenomena are seen in quantum chaotic Hermitian Hamiltonians. 

Non-Hermitian Hamiltonians possess strikingly different features from their Hermitian counterparts. The developing field of non-Hermitian random matrix theory presents a powerful tool set, which has only just begun to be applied, for uncovering universal phenomena in the emerging field of dissipative quantum chaos. We hope to make connections to dissipative dynamics in the near future.

\textit{Acknowledgments.}---We would like to thank Paul Bourgade, Amos Chan, Kohei Kawabata, and Shinsei Ryu for useful discussions and comments.
JKF is supported by the Institute for Advanced Study and the National Science Foundation under Grant No. PHY-2207584. 

\bibliography{draft_ETH}

\end{document}